\documentclass[sn-mathphys,Numbered]{sn-jnl}

\usepackage{graphicx}%
\usepackage{multirow}%
\usepackage{amsmath,amssymb,amsfonts}%
\usepackage{amsthm}%
\usepackage{mathrsfs}%
\usepackage[title]{appendix}%
\usepackage{xcolor}%
\usepackage{textcomp}%
\usepackage{manyfoot}%
\usepackage{booktabs}%
\usepackage{algorithm}%
\usepackage{algorithmicx}%
\usepackage{algpseudocode}%
\usepackage{listings}%
\usepackage{longtable}
\usepackage{array}
\usepackage{caption}

\begin{document}

\title[Article Title]{Deep Learning Algorithms Used in Intrusion Detection Systems - A Review}



\author[1,]{\fnm{Richard} \sur{Kimanzi}}\email{richard.mutisya23@students.dkut.ac.ke}
\equalcont{These authors contributed equally to this work.}

\author[1,]{\fnm{Peter} \sur{Kimanga}}\email{peter.kimanga23@students.dkut.ac.ke}
\equalcont{These authors contributed equally to this work.}

\author[1,]{\fnm{Dedan} \sur{Cherori}}\email{dedan.cherori23@students.dkut.ac.ke}
\equalcont{These authors contributed equally to this work.}

\author*[1,]{\fnm{Patrick} \sur{K. Gikunda}}\email{patrick.gikunda@dkut.ac.ke}

\affil[1]{\orgdiv{Department of 
Computer Science}, \orgname{Dedan Kimathi University of Technology}, \orgaddress{\street{P.O Box Private Bag} -  
\postcode{10143}, 
\state{Nyeri}, \country{Kenya}}}




\abstract{The increase in network attacks has necessitated the development of robust and efficient intrusion detection systems (IDS) capable of identifying malicious activities in real-time. In the last five years, deep learning algorithms have emerged as powerful tools in this domain, offering enhanced detection capabilities compared to traditional methods. This review paper studies recent advancements in the application of deep learning techniques, including Convolutional Neural Networks (CNN), Recurrent Neural Networks (RNN), Deep Belief Networks (DBN), Deep Neural Networks (DNN), Long Short-Term Memory (LSTM), autoencoders (AE), Multi-Layer Perceptrons (MLP), Self-Normalizing Networks (SNN) and hybrid models, within network intrusion detection systems. we delve into the unique architectures, training models, and classification methodologies tailored for network traffic analysis and anomaly detection. Furthermore, we analyze the strengths and limitations of each deep learning approach in terms of detection accuracy, computational efficiency, scalability, and adaptability to evolving threats. Additionally, this paper highlights prominent datasets and benchmarking frameworks commonly utilized for evaluating the performance of deep learning-based IDS. This review will provide researchers and industry practitioners with valuable insights into the state-of-the-art deep learning algorithms for enhancing the security framework of network environments through intrusion detection.}




\maketitle

\section{Introduction}\label{sec1}


\par
Application systems and computer networks are essential for effective business processes in today's quickly evolving technology world. They enable the seamless exchange of resources, including data, processing power, storage, and information\cite{bib1}. As the need for automated systems that can quickly and efficiently accomplish organizational goals has grown, so has the use of application systems in networked environments\cite{bib2}. At the same time, privacy and security issues in application systems and networks have become more well-known, highlighting the necessity of strengthening security controls against the constantly changing threats in today's cyberspace \cite{bib3}. Within the larger framework of modern civilization, the Internet is a vital resource that promotes international communication and information sharing. The Internet is essential for exchanging important information among authorities or wirelessly transmitting basic photographs on social media \cite{bib4}. Even with significant progress, persistent problems, including vulnerability to cyberattacks and difficulties enforcing international protection rules, exist, necessitating proactive measures on the part of businesses to defend their interests from possible security breaches and incursions.\par
The need to shield networks from an array of threats, including denial-of-service (DoS) assaults, unauthorized information disclosure, and data manipulation or destruction, is more critical than ever in an era where computer systems are becoming increasingly interconnected. In the realm of modern network security, ensuring the availability, confidentiality, and integrity of vital information systems has become paramount \cite{bib5}.  When paired with firewalls, intrusion detection systems (IDS) are among the essential security elements that can effectively handle a wide range of security risks \cite{bib6}. IDS techniques fall into two categories: Network Intrusion Detection Systems based on signatures and Anomaly Detection Systems based on anomaly detection. NIDS relies on signatures to identify intrusions and does so by comparing patterns across all the data they retrieve. On the other hand, anomaly-based alerts for intrusions detect any appreciable departures from the typical traffic pattern in the user behavior being monitored. Consequently, anomaly detection NIDS performs better when dealing with new attack patterns, while signature-based NIDS has a higher detection rate for recognized attack types. However, it frequently results in false alarms because of variations in intruder behavior \cite{bib4}. Finding abnormal patterns in the system audit record might aid in identifying security breaches.\par

In the past ten years, researchers have put forth a variety of Machine Learning and Deep Learning techniques to enhance Intrusion Detection Systems' capacity to identify hostile efforts \cite{bib7}. One kind of artificial intelligence technique called machine learning has the potential to extract pertinent facts from massive databases automatically. When sufficient training data is provided, machine learning-based intrusion detection systems can achieve higher detection levels. Additionally, machine learning-based IDSs are easy to create and implement since they do not heavily rely on domain expertise \cite{bib7}. One branch of machine learning that can produce very good outcomes is called deep learning. Jakhar and Kaur define Deep Learning as a subset of Machine Learning that is applied in computational problem-solving by using models and algorithms that mimic biological neural networks of the brain. Deep Learning, just like the brain, works by interpreting information, classifying it, and assigning it to various classes \cite{bib8}. Functionally, Deep Learning incorporates a set of Machine Learning algorithms that attempt learning using artificial neural network concepts in multiple levels that correspond to different levels of abstraction \cite{bib9}.  Feature representations could be learned using deep learning techniques to transform unprocessed input into final outputs. One characteristic of deep learning that sets it apart is its deep structure, which consists of many hidden layers. However, traditional machine learning models have one or no hidden layer, including the support vector machine (SVM) and k-nearest neighbor (KNN) \cite{bib7}. These conventional machine learning models are frequently called "shallow models."\par

Deep learning approaches perform better when working with vast amounts of data than standard machine learning techniques \cite{bib6}. Nevertheless, there are many barriers that IDS systems must overcome to promptly detect malicious intrusions due to the exponential increase in network traffic and associated security risks. Deep Learning Techniques for Intrusion Detection Systems (IDS) are currently being researched, and there is still more space to advance this technology for effective intrusion detection in IDS.\par

This systematic study aims to examine the current algorithms utilized to create intrusion detection systems and explain how they have been applied to guarantee everyone's safety in cyberspace. This report will consider the current research from 2019 to 2023 to achieve this goal. The following research questions will be addressed in this work:

    \begin{enumerate}
      \item{Which Deep Learning methods have been employed in IDS in the last five years?}
      \item{What is the application area of deep learning algorithms in IDS?}
      \item{What are the attacks covered by various datasets used in developing Deep Learning Network Intrusion detection Systems?}
    \end{enumerate}

This paper is organized in the following order. Section two surveys the related literature works regarding the deep learning algorithms used in Intrusion Detection Systems. Section three reports the results of performance for the various algorithms. Section four gives a conclusion to this paper as well as giving directions for future works.


\section{Related work}\label{sec3}
This section will analyze the various deep learning algorithms based on recent research within the last five years. But first, we define Deep Learning as a computing concept in Machine Learning and Artificial Intelligence. 
Deep learning models that have been applied in Intrusion Detection Systems include Deep Belief Networks (DBNs), Convolutional Neural networks (CNNs), Recurrent Neural Networks (RNNs), Long-Short-Term-Memory (LSTM), Multilayer Perceptrons (MLPs), autoencoders, Self Organizing Maps (SOMs) \cite{bib10}.

\subsection{Convolutional Neural Networks (CNNs)}
Khan et al. in \cite{bib11} proposes an Improved CNN Model for network intrusion detection that combines the CNN algorithm and softmax algorithm. This model executes three processes in its intrusion detection method. First, data pre-processing is carried out through the conversion of symbolic data to numeric data then normalization is of the data undertaken. Secondly, a CNN model with three hidden layers is used for data training and feature extraction. Finally, the softmax classifier is utilized when obtaining the classification results. The proposed NIDS is tested using the KDD99 data set. The results of the experiment exhibit an improved performance in terms of the accuracy in intrusion detection compared to Support Vector Machines and Deep Belief Networks. This accuracy is improved by increasing the number of epochs in the proposed model. The proposed model should be tested for its performance against newer state-of-the-art network intrusion algorithms and its applicability in IoT and Cloud computing where there are large-scale data and sophisticated attacks.\par
A novel NIDS is proposed by Liu et al. in \cite{bib12} that improves intrusion detection by employing CNN combined with Fast Fourier transformation for representation learning. The proposed method works by converting network traffic to an image after the preprocessing stage by modeling a visual data transformation representation algorithm. A CNN classifier i.e. the ResNet50 framework by MathWorks, is then applied for image processing. Image processing for anomaly detection is the core method that Liu et al. propose to improve intrusion detection using CNN. The algorithm is tested using the KDDCup’99 dataset. The proposed algorithm shows overall better performance in both binary and multi-class classification. It also shows improved accuracy when compared with other algorithms. However, this algorithm does not detect U2R and R2L anomalies since they are very few in the used data set. Furthermore, it is not tested against state-of-the-art deep learning algorithms to compare its performance.\par
Hu et al., built and implemented a Wi-Fi sensing system for detecting intrusion employing Channel State Information (CSI) at the physical layer of a Wi-Fi network as a detection signal in \cite{Hu2021}. To boost the sensitivity of a passive intrusion detection system, they used a path decomposition method and Deep learning convolutional neural network. CNN was used to enable the computer to learn and identify intrusion without extracting numerical characteristics. They used Channel State Information (CSI) dataset. In four different cases or scenarios that they had in their experiment, the average detection accuracy was 98.69\% with a single person and 98.91\% with multiple participants. They concluded that in comparison to prior methods, IDSDL can detect human movements on non-line of sight (NLOS) pathways more sensitively and increase system reliability. \par
Arun et al. \cite{Arun2023} introduced an innovative intrusion detection algorithm designed to bolster network security in the wake of escalating thefts and system breaches. Leveraging Convolutional Neural Network (CNN), the authors present a high-level Intrusion Detection System (IDS) that meticulously examines network abnormalities, representing a substantial advancement in information security technology. The study emphasizes the pressing need for resilient systems capable of identifying novel attacks amidst the rapid evolution of networks and technology. By amalgamating machine learning and deep learning algorithms, the proposed IDS focuses on discerning elusive assault bundles. Significantly, the advanced system demonstrated a noteworthy enhancement in precision, fortifying its ability to accurately detect intricate attack patterns. This study underscores the pivotal role of Intrusion Detection Systems in fortifying information security and addressing the surge in cyber threats within the dynamic technological landscape (citation28). The specific network intrusion detection techniques and datasets employed in this study is unsw-nb15
Arun et al. \cite{Arun2023} study the escalating occurrences of thefts and system breaches that underscore the pressing need for advanced security measures. They introduce an innovative approach to address this challenge through the deployment of a high-level Intrusion Detection System (IDS) incorporating Convolutional Neural Network (CNN) technology. By leveraging machine learning and deep learning algorithms, the IDS aims to detect network abnormalities effectively. The primary objective of the study is to identify elusive attack bundles, enhancing precision in intrusion detection. The proposed system exhibits remarkable precision improvements, indicating its efficacy in recognizing intricate attack patterns. The research emphasizes the significance of information security technology, particularly IDS, in the evolving landscape of network threats. Through the application of high-performance intrusion detection systems, the study demonstrates a proactive approach to identifying and mitigating potential security breaches, thereby contributing to the overall enhancement of network security.\par
Alissa et al. in \cite{Khalid2022} address the pressing challenge of cybersecurity in the context of the Internet of Things (IoT), where a multitude of small smart devices transmit vast amounts of data over the Internet. Recognizing the inherent security flaws in these IoT gadgets, exacerbated by the lack of hardware security support, the study focuses on the development of an innovative solution for lightweight intrusion detection in resource-constrained IoT environments. The proposed approach, named Planet Optimization with Deep Convolutional Neural Network for Lightweight Intrusion Detection (PODCNN-LWID), leverages a Deep Convolutional Neural Network (DCNN) for intrusion identification and integrates Planet Optimization (PO) as a hyperparameter tuning process. The primary goal of PODCNN-LWID is to identify and categorize intrusions efficiently. The model undergoes two major processes: intrusion identification using DCNN and hyperparameter tuning through PO. Experimental validation on a benchmark dataset demonstrates the effectiveness of the PODCNN-LWID model, showcasing enhancements over alternative approaches in terms of intrusion detection performance. The study contributes valuable insights into mitigating cybersecurity challenges in resource-constrained IoT environments.\par
Man and Sun in \cite{Manandsun2021} proposed a CNN model for network intrusion detection strategy based on residual learning and focal loss. It aimed to enhance network security by introducing a novel approach based on residual learning. The techniques employed involved the implementation of residual connections within the neural network architecture using UNSW-NB15 dataset.  Its results revealed that residual learning models are easier to train, with smaller loss values on training data and higher accuracy rates on testing data.

\subsection{Deep Belief Networks (DBNs)}
A fuzzy aggregation method for intrusion detection is proposed by Yang et al \cite{bib14}. This system is applicable in the Internet of Things (IoT) and cloud environments as well as software-defined networks. This research tries to address the problem of large data sets in the mentioned environments as well as capturing emerging attacks that are not captured in the training data set by combining the Modified Density Peak Clustering Algorithm (MDPCA)and the Deep Belief Network (DBN) model. MDPCA is used in the extraction of similar features from intricate and extensive network data then clustering it into various training subsets based on the similarity of the network traffic. This reduces the complexity of the training subsets. DBN then automatically applies the feature extraction and classification module to the training subsets where extraction of sophisticated abstract features is carried out while avoiding manual procedures and heuristic rules. The experimental tests of this proposed IDS are carried out using the NSL-KDD and UNSW-NB15 data sets. The results show overall best performance when compared to other state-of-the-art models in terms of accuracy and detection rate on the UNSW-NB15 except for CASCADE-ANN which ranks higher than MDPCA-DBN on the False Positive Rate parameter. Further tests show overall best performance when compared to other state-of-the-art models in terms of accuracy, detection rate, and false positive rate on the NSL-KDD data set. MDPCA-DBN also applies at least two test data sets which produce almost similar results in terms of efficiency thus proving its efficacy in solving the stated multi-class classification problems. The areas of application are clearly stated as well ranging from Software-defined networks, to IoT and Cloud Environments.\par 
Thamilarasu and Chawla in \cite{bib15} identified a security vulnerability in IoT due to their large attack surface. They propose a secure, portable, and ready-to-deploy IDS suitable for an IoT environment. The heterogeneous nature of IoT networks makes them susceptible to cyber-attacks\cite{bib16}. The proposed algorithm forms a DNN using a DBN. The ability of DBN to be trained using unsupervised learning allows the DNN to be trained using unsupervised learning which is faster compared to supervised learning. Ideally, network virtualization and a DNN binary classifier are employed in detecting abnormal traffic in an IoT network. Upon testing the proposed model, there is overall improved accuracy over Inverse Weight Clustering in the simulation environment and the testbed. The wormhole attack however performs poorly in the test bed environment as compared to the simulation environment. This model is one of the few that have been tested in a testbed with real IoT network traffic with fairly good performance. However, the study does not state the dataset used in the simulation environment. 
\subsection{Autoencoders (AE)}

Sandeep Gurung et al. in \cite{bib17} propose a deep learning intrusion detection system that uses a sparse auto-encoder and logistic regression in solving the network intrusion problem. A sparse auto-encoder that has a sparsity constraint is used to train the model. Logistic regression is then applied in the classification phase that results to a binary classification of either a normal user or intruder.  The model is trained and tested using the NSL-KDD data set. This model results to a higher accuracy rate when compared to Signature-Based Intrusion Detection approaches with reduced chances of False Positives and False Negatives. This approach is best applied in real-time network monitoring servers although there is no evidence of real-world application of this model.\par
Telikani and Gandomi in \cite{TELIKANI2021100122} propose a Cost-sensitive stacked auto-encoders model for intrusion detection in the Internet of Things networks with the aim of solving the class imbalance problem in IDSs. This model works by modifying the cost function of a Stacked Auto-Encoder to be more sensitive toward misclassification of minority class by reducing the effect of unbalanced data on the performance of intrusion detection systems. Batch Normalization method is used to speed up the learning phase while Softmax classifier, a multi-class version of Logistic Regression, is used in the classification of anomalies. Extraction of probabilities for each class is achieved using cross-entropy loss. This study uses the  KDD cup 99 and NSL-KDD Datasets to train at test the proposed model. Experimental results show that it performs better than Stacked Auto-Encoder (SAE) and Non-symmetric Deep Auto-Encoder (NDAE) in all metrics when tested using both datasets.\par
An IoT intrusion detection system using asymmetric parallel auto-encoder (APAE) is proposed by Basati and Faghih in \cite{Basati2023} for detecting real-time cyber-attacks in IoT networks. This study uses the UNSW-NB15, CICIDS2017, and KDDCup99 datasets for training, testing and validating the model. An APAE is first trained on a dataset in the proposed model in order to estimate the identity function for the training data. In order to estimate the identity function for the training data, an APAE is first trained on a dataset in the suggested model. After that, the finished model is achieved by joining the first three components of the learned APAE—the transfer layer, encoder, and latent features—with a fully linked classifier layer at the end. Subsequently, the final model is retrained using the training data in order to determine the classifier weights and adjust the APAE encoder weights for precise classification. The proposed APAE model Slightly better in classification compared to other proposed auto encoder models. It also performs the best classification in the minority classes. This study achieves a lightweight NIDS suitable for IoT networks and devices which are limited in terms of processing power, memory, and energy efficiency.\par
A study by Kamalov et al. in \cite{kamalov}, the authors present an Autoencoder-based Intrusion Detection System (IDS) designed for detecting Distributed Denial of Service (DDoS) attacks. The algorithm identifies intrusions by flagging anomalous traffic flows with higher reconstruction loss. The study utilizes the CSE-CIC-IDS2018 dataset for evaluation. The proposed strategy demonstrates superior performance compared to benchmark unsupervised systems in the detection of DDoS attacks.

\subsection{Deep Neural Networks (DNNs)}

A Scalable Hybrid Intrusion Detection System Alertnet (SHIA) is proposed by R. Vinayakumar et al. in \cite{bib18}. This approach Employs distributed deep learning with DNNs for analyzing and handling large sets of data in real time with the aim of solving the problem of detection and classification of unforeseen and unpredictable cyber-attacks. This intrusion detection system is modeled using a scalable computing framework based on Apache Spark Cluster Computing platform. A System Call trace is used for text representation and classification of process behaviours. The DNN is modelled using Feed forward Neural (FFN)Network and Multilayer Perceptron (MLP). This paper trains and tests the model with several datasets including KDD Cup 99, NSL-KDD, UNSW-NB15, WSN-DS, CICIDS 2017, Kyoto. The test results show overall superior performance than classical machine learning classifier models for NIDs and HIDs such RF, SVM, KNN, LR, AB. The experiments using a variety of datasets shows the versatility of the model and its ability to work in an environment with large scale data. The authors offer the proposed model for use Efficient real-time network monitoring for the network traffic and host-level operation such as Malicious host detection.

A hybrid classifier model for intrusion detection known as Spider Monkey Optimization and Deep Neural Network (SMO-DNN) is proposed by Khare et al. in \cite{bib20}. In this study the NSL-KDD and KDD Cup 99 Datasets have been used. At the data preprocessing stage, data cleansing is done using the Min-Max normalization technique then 1-N encoding is used to achieve consistent and unified data. The Spider Monkey Optimization is utilized for dimensionality reduction after which the reduced dataset is split into the Training and testing datasets then forwarded to the Deep Neural Network for classification which uses the the softmax classifier at the output layer. The resulting classification is either normal traffic or intrusion traffic. SMO-DNN is tested in comparison with Principal Component Analysis (PCA) PCA-DNN and DNN and it exhibits better performance than the two models. PCA is tested as a reduction algorithm with DNN as the classifier. This model is a binary classifier that is applicable in large network. However, the study should have tested the model against other state-of-the-art network intrusion detection algorithms to further prove its efficacy.\par
A Transfer Learning for Network Intrusion Detection System (TL-NID) is proposed by Masum and Shahriar in \cite{masum2021}. This model combines a Deep Neural Network for Classification while transfer learning is used for feature extraction. This algorithm executes two procedures in finding a solution to the malicious network intrusion problem. Two step method is modelled using the pre-trained VGG-16 for feature extraction. Extracted features are then fed into a DNN for classification. The sigmoid layer of the DNN carries out binary classification. This model is trained and tested using the NSL-KDD dataset. The model is tested against traditional machine learning models i.e. Support Vector Machines (SVM), Logistic Regression (LR), Decision Tree (DT), and Random Forest (RF) and its performance is analyzed. The proposed model exhibits the best Performance in terms of Accuracy, Recall and F1-Score using both KDDTest+ and KDDTest-21 datasets. Decision trees (DT) perform better than the proposed model on Precision and False Alarm Rate using the KDDTest+ dataset. SVM performs better than the proposed model on Precision and False alarm rate using the KDDTest-21 dataset.

Khan et al., in \cite{Khanetal2021} suggested a DNN-based intrusion detection system for MQTT-enabled IoT smart systems. The suggested model's performance was evaluated using a MQTT-IoT-IDS2020 dataset and another Message Queuing Telemetry Transport (MQTT) dataset. The proposed DL-based IDS's performance with a default learning rate and the ADAM optimizer was compared to the performance of conventional ML-based IDSs such as KNN, NB, DT, and RF. The suggested model was further evaluated for binary-class and multi-class attack classification with different activation functions at the output layers. According to the results, the DL-based model for Bi-flow and Uni-flow featured data reached 99\% and 98\% accuracy for binary and multi-class attack classification, respectively.\par
Gulia et al. in\cite{Gulia2023} have made noteworthy strides in advancing network security, specifically within the domain of cloud computing, by introducing an Intrusion Detection System (IDS) that leverages the Group-Artificial Bee Colony Algorithm (G-ABC) algorithm alongside a Deep Neural Network (DNN). In their study, this innovative approach effectively addressed security concerns inherent in cloud computing environments, resulting in a commendable 96\% detection rate for improved intrusion detection capabilities. Beyond the proposed IDS, the authors conducted an exhaustive literature review, highlighting various studies on both host-based (HIDS) and network-based (NIDS) IDSs, showcasing diverse methodologies, including hybrid models, optimization techniques, and robust IDS designs. These collective efforts contribute to overcoming challenges such as system complexity and limitations in alert systems. The proposed two-phase framework, combining G-ABC for optimal feature selection and DNN for classification, not only aims to elevate IDS performance but also demonstrates promising results, solidifying its potential impact on the future of network security in cloud environments.\par
Syariful et al. \cite{Syariful2022} delve into the critical domain of cybersecurity within the context of the Internet of Things (IoT). As the proliferation of IoT devices continues to enhance human lives, it also amplifies the susceptibility of networks to potential attacks. Recognizing the inherent vulnerabilities in IoT networks, the study employs artificial intelligence as a proactive measure to safeguard these networks by detecting and preventing attacks effectively. The research specifically focuses on network detection utilizing the Deep Neural Network (DNN) algorithm, a sophisticated artificial intelligence technique. The experimentation involved testing on the UNSW Bot-IoT dataset, with training data comprising 75\% of the overall dataset. Remarkably, the DNN algorithm demonstrated exceptional performance, achieving an average detection accuracy of 99.999\%. Despite the presence of a small validation loss indicative of overfitting, the study attests to the robust capability of the algorithm in effectively identifying and mitigating attacks in IoT networks.\par

\subsection{Recurrent Neural Networks (RNNs)}

Ashwaq et al., \cite{Ashwaq2022} address the escalating vulnerabilities and attacks prevalent in the Internet of Things (IoT) environment. As the IoT landscape continues to expand, facilitating the connection of over 20 billion items by 2024, the imperative of securing this interconnected ecosystem becomes increasingly evident. To address this, the researchers propose a novel approach employing Recurrent Neural Network (RNN) deep learning algorithms for intrusion detection within the IoT environment. The study utilizes the NSL-KDD dataset for both training and testing purposes. The proposed RNN model demonstrates notable success, achieving an accuracy of 87\% in detecting intrusions. The research underscores the significance of leveraging machine learning and deep learning techniques for enhancing security within the evolving IoT paradigm. Furthermore, the authors express their commitment to future work, intending to explore optimization algorithms to further augment the detection accuracy of their proposed model.

\subsection{Self-Normalizing Neural Networks (SNNs)}

A novel network-based detection system for IoT attacks is modelled by Aldhaheri et al. in \cite{aldhaheri_deepdca_2020}. The BoT-IoT dataset is used with the proposed algorithm combining Self-Normalizing Neural Networks (SNNs) with Dendritic Cell Algorithm (DCA). The aim is to classify IoT intrusion and minimizing false alarm generation. The proposed algorithm automates and smoothens feature extraction and categorization stage which in turn improves the classification performance. Features are either classified as safe or dangerous signals using the Self-Normalizing Neural Network (SNN). The proposed DeepDCA model exhibits better performance as a classifier compared to other models such as KNN, NB, MLP and SVM. This model also showed a high detection rate for IoT attacks with a high accuracy rate and a low False-Positive Rate. Self-normalization Neural Network is a fairly new technique and further studies could be carried out to explore its potential in solving intrusion detection problems and other deep learning problem domains.

\subsection{Multi-Layer Perceptron (MLP)}

Louati and Ktata propose a deep learning based multi agent system for intrusion detection (DL-MFID) in \cite{bib39}. Data pre-processing is carried out using pre-processor agent to achieve numericalization and normalized data. Feature reduction is then performed using an auto-encoder as a reducer agent to select relevant features. Finally multi-class classification is done using into 5 classes using two classifier agents i.e. MLP and KNN. This model is trained and tested using the KDD cup 99 dataset. Experimental results show that this model exhibits the highest accuracy rate of 99.95\% compared to other proposed models which used the same dataset. This model Has however not been applied in real network traffic to test its efficiency. The approach of using multiagent classifiers is a new approach that requires further research to study how it can improve network intrusion detection systems. \par
Abdulrahman et al. in \cite{mohammed2020multilayer} addressed the critical issue of securing network systems against Distributed Denial of Service (DDoS) attacks by developing two effective Intrusion Detection System (IDS) models based on multilayer perceptron neural networks (MLP). In the first model, MLP1, a one-hidden-layer MLP with 38 input nodes, 11 hidden nodes, and 5 output nodes achieved a remarkable detection accuracy of 95.6\%. Building upon this, the second model, MLP2, enhanced with two hidden layers, surpassed MLP1, achieving an accuracy 2.2\% higher. The study demonstrates the efficacy of MLP-based IDS in accurately classifying different request types and highlights the importance of robust models in countering DDoS attacks. The training and evaluation were conducted using the NSL-KDD dataset, showcasing the practical application of MLP-based IDS in real-world scenarios\par
A multilayer perceptron for network intrusion detection is proposed by Rosay et al. in \cite{Rosay2022} aimed at preventing cyber-attacks in Vehicular IoT networks. This model work by training a multilayer perceptron model that quantifies the benefits of using deep learning strategies in network intrusion detection systems. The CIC-IDS2017 is used to define the IDS solution while the CSE-CIC-IDS2018 dataset is used to validate and test the proposed model. At the pre-processing stage the training Dataset is cleaned up to remove imbalances before the dataset is split to three categories namely; Training set – 50\%, validation set - 25\%, test set - 25\%. Feature selection is then carried out to remove irrelevant features before the data normalization is done using the Z-normalization technique. At the experimental level, the model is built and trained using python and the tensor flow framework as the deep learning model. Experimental results show that the proposed model performs better than traditional ML models in terms of accuracy with significantly low false positive rates. This model was further deployed in a practical system on chip vehicle allowing test to be carried out in an embedded system. Several shortcomings were identified where some attacks are underrepresented and some features are not well calculated in both datasets used by the model. Testing this model in a real-world scenario means it can be easily re-engineered for practical implementation in the automotive industry. 

\subsection{Hybrid Algorithms}
Susilo and Sari in \cite{bib13} evaluate the effectiveness of Random forests, CNN and Multi-layer Perceptrons in IoT Network Intrusion Detection Systems. Their study uses the BoT-IoT dataset to for evaluation of intrusion detection. Owing to the heterogeneous nature of IoT networks and traffic, multi-class classification of intrusions was preferred. Experimental results show that Random Forests and CNNs perform better than MLP in terms of accuracy and Area Under Curve parameters in a multi-class classification scenario. The study proposes a combination or hybridization of algorithms as future studies in implementing NIDs for IoT environment. It is worth noting that the dataset used is specific to the research problem of intrusion detection in IoT application. However, the study ought to have considered other datasets in testing the efficacy of the designed models as well as testing them against other state-of-the-art algorithms for Network Intrusion Detection.\par
Dutta et al. in \cite{bib19} propose a model for efficient network anomaly detection or cyber-attack detection system that employs a stacked ensemble classifier approach. This approach is meant for implementation in an IoT and cloud computing environment. This study used heterogenous datasets from IoT environments such as IoT-23, LITNET-2020 and NetML-2020.  Dutta et al. combines DNN with LSTM in a stack generalization approach for initial classification, then uses logistic regression as a meta classifier. The method begins with data preprocessing followed by feature engineering that uses a Deep Sparse Auto-Encoder. Classification as the third step employs DNN and LSTM models followed by logistic regression which helps in eliminating biases from training sets. The end result is a binary classification results of the data as either normal or anomalous. The experimental results shows that the proposed stacked ensemble algorithm performs better than state-of-the-art classifiers and meta-classifiers. The use of three recent datasets that fit the problem domain is an indicator that the model would perform well if implemented in a real-world scenario.\par 
In their work titled "Intrusion Detection System using MLP and Chaotic Neural Networks", Shettar et al. \cite{Pooja2021} propose a hybrid intrusion detection model that uses multi-layer perceptron and chaotic neural network to achieve its objective. They addressed the challenges in developing a robust Network Intrusion Detection System (NIDS) capable of identifying unexpected and impulsive cyber-attacks. While existing techniques focus on enhancing accuracy, the authors introduced a hybrid model integrating Multilayer Perception (MLP) and chaotic neural networks to simultaneously improve accuracy, precision, and notably, the false alarm rate—a critical aspect often overlooked in NIDS development. The methodology involved employing MLP for attack detection and chaotic neural networks to specifically target and reduce false alarm rates. The researchers conducted experiments using the KDD Cup'99 benchmark dataset, a widely used standard for intrusion detection evaluations. The results demonstrated that the hybrid approach outperformed the standalone MLP model by effectively mitigating false alarm rates while maintaining comparable accuracy, precision, and recall performance. This research contributes to the field of intrusion detection by offering a holistic solution that not only enhances the system's ability to detect cyber-attacks accurately but also addresses the vital issue of minimizing false alarms. The findings suggest that incorporating chaotic neural networks in conjunction with MLP can lead to more reliable and efficient NIDS, crucial for maintaining the integrity and security of computer networks in the face of evolving cyber threats.\par
Alkahtani and Aldhyani in \cite{Alkahtani2021} developed a comprehensive framework system for detecting intrusions based on the IoT environment. The proposed technique was developed using an IoT ID20dataset attack; a dataset from the IoT infrastructure. To identify the incursion in this framework, three advanced deep learning algorithms were used: a convolution neural network (CNN), a long short-term memory (LSTM), and a hybrid convolution neural network with the long short-term memory (CNN-LSTM) model. The dimensionality of the network dataset was reduced, and the particle swarm optimization method (PSO) was utilized to identify key characteristics from the network dataset to improve the suggested system. Deep learning methods were used to process the collected features. The results revealed that the proposed systems obtained the following levels of accuracy: CNN (96.60\%), LSTM (99.82\%), and CNN-LSTM (98.80\%).\par
Ashiku and Dagli in \cite{ASHIKU2021239} researched the use of deep learning architectures in the development of an adaptive and resilient network intrusion detection system (IDS) to identify and categorize network threats. The emphasis was on CNN with regularized multi-layer perceptron to enable customizable IDS with learning power to identify known and unknown network threats. They used UNSW-NB15 dataset. When compared to the results of similar deep learning-based network IDSs, the suggested deep-learning classification architecture combined with the semi-dynamic hyperparameter tuning approach revealed considerable improvements to multiclass models.  According to the models, their proposed approach achieved an overall accuracy of 95.4\% and 95.6\% for pre partitioned and user defined multiclass categorization, respectively.\par
Muhuri et al in \cite{bib21} propose a network intrusion detection system based on RNN model. They combine Long Short-Term Memory and RNN for classification while Genetic Algorithm is utilized for optimal feature selection. This model is aimed at ensuring high detection rates compared to existing models using the NSL-KDD dataset. Optimal feature selection using GA is applied to the LSTM-RNN model to remove bias in the classification process. This helps in minimizing misclassification rate, training time as well as maximizing the accuracy of the model. In the proposed model the original features in the dataset are 122 and thus GA optimizes these features to a subset of 99 features. The proposed model LSTM-RNN with GA produced the best performance compared to RF and SVM in both binary and multiclass classification tests. The LSTM-RNN with GA therefore offers significant improvement in NIDs accuracy and detection rates compared to traditional machine learning approaches. This model is applicable in large dataset environments such a large computer networks with high traffic.\par
In their quest to fortify the security of the Internet of Things (IoT), Safi et al. \cite{Safi2022} address the escalating concern of cyberattacks targeting IoT networks. Recognizing the susceptibility of numerous IoT nodes that process sensitive data, the researchers propose a novel Intrusion Detection System (IDS) leveraging a deep-convolutional-neural-network (DCNN). The DCNN architecture encompasses two convolutional layers and three fully connected dense layers, aiming to enhance overall performance while minimizing computational demands. The study conducts rigorous experiments on the IoTID20 dataset, employing diverse optimization techniques, including Adam, AdaMax, and Nadam. Performance evaluation metrics such as accuracy, precision, recall, and F1-score are systematically analyzed. Notably, the proposed DCNN-based IDS exhibits superior accuracy and robustness when compared to existing deep learning (DL) and traditional machine learning (ML) techniques. This research contributes to advancing intrusion detection methodologies tailored for IoT environments, providing an effective shield against evolving cyber threats.\par
S Amutha et al. \cite{samutha2022} proposed an innovative approach, the NID-Recurrent Neural Network (RNN), to enhance the efficiency of Secure Network Intrusion Detection Systems (NIDS). Recognizing the importance of preventing malicious activities and cyber-attacks, the study delves into the realm of deep learning, specifically focusing on the integration of NID with Long Short-Term Memory (LSTM). Unlike traditional machine learning methods that rely on manually crafted features, deep learning approaches, including deep neural network (DNN), Convolutional Neural Network (CNN), and RNN, are explored for their potential to elevate the performance of intrusion detection. The researchers conduct a comprehensive analysis, comparing the accuracy and precision of each model, considering binary and multi-class classifications on the NSL-KDD dataset. Notably, the results reveal an 8\% improvement in accuracy for the proposed NID-RNN approach, outperforming other deep learning algorithms. The RNN model achieves an impressive 99.4\% accuracy in classifying attack types, showcasing its efficacy in bolstering network security.\par
Sunil et al. In \cite{Gautam2022}this research on intrusion detection systems (IDS), the authors introduced a novel approach by developing a composite model that combines Bidirectional Recurrent Neural Network (RNN) using Long Short-Term Memory (LSTM) and Gated Recurrent Unit (GRU). The study utilized the CICIDS2017 dataset for simulations and evaluation. Notably, the hybrid RNN IDS achieved an outstanding 99.13\% classification accuracy in predicting network attacks, showcasing the effectiveness of the proposed model. This performance surpassed the accuracy and False Positive rate of the Naïve Bayes classifier, highlighting the superiority of the deep learning-based hybrid approach. The model's efficiency was further emphasized by its ability to achieve robust results using only 58\% of the dataset attributes, demonstrating resource efficiency and practical applicability.\par
Sydney Mambwe Kasongo in \cite{KASONGO2023}, the escalating volume of data transmission in communication infrastructures prompted the implementation of an Intrusion Detection System (IDS) framework utilizing advanced Machine Learning (ML) techniques. The framework employed various Recurrent Neural Networks (RNNs), such as Long-Short Term Memory (LSTM), Gated Recurrent Unit (GRU), and Simple RNN, to enhance the security of network systems. To address the challenge of low test accuracy in detecting new attacks with increasing feature dimensions, an XGBoost-based feature selection algorithm was integrated into the framework. The evaluation was performed on benchmark datasets, including NSL-KDD and UNSW-NB15, revealing significant achievements. For binary classification tasks on NSL-KDD, the XGBoost-LSTM model demonstrated superior performance with a test accuracy of 88.13\%, a validation accuracy of 99.49\%, and a training time of 225.46 seconds. On UNSW-NB15, the XGBoost-Simple-RNN emerged as the most efficient model with a test accuracy of 87.07\%. For multiclass classification, the XGBoost-LSTM achieved a test accuracy of 86.93\% on NSL-KDD, while the XGBoost-GRU obtained 78.40\% on UNSW-NB15. These outcomes underscore the effectiveness of the proposed IDS framework in comparison to existing methods, offering optimal intrusion detection performance in the face of evolving cyber threats.

\section{Summary of discussed algorithms}\label{sec4}

Table \ref{tab1} summarizes the methods by each of the intrusion detection mechanisms discussed in section \ref{sec3} of literature.

\begin{longtable}
{p{1.9cm}p{3cm}p{3.7cm}p{4cm}}
\captionsetup{justification=raggedright,singlelinecheck=false}
\caption{Summary of reviewed Deep Learning Algorithms and methods used in their implementation}\label{tab1}%
\\
    \hline
    \textbf{Reference} & \textbf{DL Algorithm} & \textbf{Problem} & \textbf{Method} \\
    \midrule
    \endhead 
    \cite{bib11} & CNN  & Low accuracy in intrusion detection.  & Combine convolution and pooling operations. \\
    \midrule
    \cite{bib12}    & CNN  & Low classification accuracy  & Conversion of network traffic to an image by modeling a visual data transformation Representation algorithm\\
    \midrule
    \cite{Hu2021}    & CNN   & Passive intrusion detection  & Path decomposition method and CNN for intrusion detection\\
    \midrule
    \cite{Arun2023}    & CNN   & IDS for system Breaches  & convolution neural network for anomaly detection techniques \\
    \midrule
    \cite{Khalid2022}    & Deep CNN   & IDS for resource-constrained IoT  & Deep CNN as intrusion classifier, Planet algorithm as hyperparameter optimizer \\
    \midrule
    \cite{Manandsun2021}    & CNN   & Efficiency in network intrusion detection  & Residual learning and Focal Loss \\
    \midrule
    \cite{bib14}    & DBN   & IDS for cloud \& IoT environment. Emerging attacks don’t appear in the training datasets  & Fuzzy aggregation approach \\
    \midrule
    \cite{bib15}    & DBN   & IoT Security networks due to their large attack surface  & Categorizing network traffic into sessions \& investigating anomalous characteristics of network activity \\
    \midrule
    \cite{bib17} & Sparse Auto-Encoder(SAE) & Real-time network monitoring and detection of new attacks  & SAE for Unsupervised Feature learning \& logistic regression classifier \\
    \midrule
    \cite{Basati2023} & APAE & Detecting real-time cyber-attacks in IoT networks & Utilize dilated and standard convolutional filters to extract both locally and long-range information around individual values in the feature vector \\
     \midrule
    \cite{kamalov} & Auto-Encoder & Network intrusion detection & Identification of intrusions by flagging anomalous traffic flows with higher reconstruction loss \\
     \midrule
    \cite{bib18} & DNN & Detection and classification of unforeseen and unpredictable cyber-attacks & DNN that employs Feed forward Neural (FFN)Network and Multilayer Perceptron (MLP) \\
     \midrule
    \cite{bib20} & DNN \& SMO & Network Intrusion Detection & min-max normalization technique, 1-N encoding, SMO used to reduce dimensionality of the dataset \& DNN for binary classification. \\
     \midrule
    \cite{masum2021} & DNN with Transfer learning & Network Intrusion Detection & Pre-trained VGG-16 used for feature extraction \& Extracted features are fed into a DNN for classification Extracted \\
     \midrule
   \cite{Khanetal2021} & DNN & Intrusion detection in MQTT-enabled IoT smart systems & Deep Learning model IDS with default learning rate and ADAM optimizer \\
     \midrule
    \cite{Gulia2023} & DNN, G-ABC & Intrusion detection in cloud computing environment & G-ABC for optimal feature selection, DNN as a classifier \\
     \midrule
     \cite{Syariful2022} & DNN & Intrusion detection in IoT environment & One Dimension vector, feature reduction, Z-Score normalization \& DNN Classifier \\
    \midrule
    \cite{Ashwaq2022} & RNN & Intrusion detection in IoT environment & Dataset splitting, Scalar-based normalization, feature extraction, classification using RNN \\
    \midrule
    \cite{aldhaheri_deepdca_2020} & SNN \& Deep DCA & Intrusion detection for IoT environment & Self-normalizing Neural Network for feature selection \& categorization, Dendritic Cell Algorithm for classiffcation  \\
    \midrule
    \cite{bib39} & MLP, KNN, Auto-Encoder & Network intrusion detection & Deep learning multi-agent system approach. Auto-encoder for feature reduction, MLP \& KNN as Classifiers \\
    \midrule
    \cite{mohammed2020multilayer} & MLP & Accuracy in Network intrusion detection & One hidden layer MLP, Two hidden layer MLP as classifiers, Dataset trained with 100 epochs \\
    \midrule
   \cite{Rosay2022} & MLP & Prevention of cyber-attacks in vehicular IoT & Dataset cleaned up, split dataset to test, validation and training sets, feature selection, Z-normalization, training model  \\
    \midrule
    \cite{bib13} & CNN \& MLP & IDS for heterogeneious IoT networks & Deep learning implementation method not clear. Python and python frameworks i.e. numpy, pandas, keras and tensor flow used in experiments. \\
     \midrule
    \cite{bib19} & DNN, LSTM \& Logistic regression & Anomaly detection in IoT and Cloud environments & combination of DNN with LSTM in stack generalization approach for initial classification. logistic regression as a meta classifier \\
     \midrule
    \cite{Pooja2021} & MLP \& Chaotic Neural Network &  Robust network intrusion detection system & MLP for attack detection and Chaotic Neural Network for reduction of false alarm rates \\
     \midrule
    \cite{Alkahtani2021} & CNN, LSTM \& PSO & Intrusion detection in IoT environment & PSO for data pre-processing, Hybrid CNN and LSTM used for feature selection and classification. \\
     \midrule
    \cite{ASHIKU2021239} & CNN \& MLP & Network intrusion detection & CNN with regularized MLP, semi-dynamic hyperparameter tuning. \\
     \midrule
    \cite{bib21} & LSTM-RNN, Genetic Algorithm (GA) & Optimal attack classification for NIDS & LSTM-RNN Classifier with GA feature selection Architecture \\
     \midrule
    \cite{Safi2022} & LSTM \& Gated Recurrent Unit (GRU) & DDoS in IoT & Eliminating the vanishing gradient problem using LSTM and GRU. \\
    \midrule
    \cite{samutha2022} & RNN \& LSTM & Network intrusion detection & RNN-LSTM trained model using UNSW-NB18 dataset. \\
    \midrule
    \cite{Gautam2022} & RNN, LSTM \& GRU & Network intrusion detection & Feature selection and bidirectional RNN. LSTM sequence for forward pass \& GRU sequence for backward pass\\
    \midrule
    \cite{KASONGO2023} & RNN, LSTM \& GRU & Network intrusion detection & initial dataset normalization, Feature selection using XGBOOST, Training, validation and testing using Deep Learning algorithms i.e. RNN, LSTM and GRU. \\
    \hline
\end{longtable}

\begin{longtable}
{p{1.9cm}p{2.5 cm}p{5 cm} }
\captionsetup{justification=raggedright,singlelinecheck=false}
\caption{Summary of reviewed Papers and the attacks discussed }%
\\
    \hline
    \textbf{Reference} & \textbf{Dataset} & \textbf{Attacks}  \\
    \midrule
    \endhead 
    \cite{bib11} & KDD99 datasets  & DOS, R2L, U2R, and probe.  \\
    \midrule
    \cite{bib12}    & NSL-KDD  & DoS, Probe, R2L, U2R.\\
    \midrule
    \cite{Hu2021}    & CSI Data   & Motion Detection\\
    \midrule
    \cite{Arun2023}    & UNSW-NB15   & DoS, Probe, R2L, U2R.  \\
    \midrule
    \cite{Khalid2022}    & CICIDS2017 dataset   & Botnet, DoSSlowhttptest, FTP-Patator, SSH-Patator, DoSGoldenEye, DoSslowloris, Heartbleed, PortScan, DDoS, DoSHulk \\
    \midrule
    \cite{Manandsun2021}    & UNSW-NB15 dataset   & Analysis, Backdoors, DoS, Exploits, Fuzzers, Generic, Recon, Shellcode, Worms \\
    \midrule
    \cite{bib14}    & NSL-KDD   & DoS, Probe, R2L, U2R \\ \\
    
       & UNSW-NB15 datasets   & Analysis, Backdoors, DoS, Exploits, Fuzzers, Generic, Recon, Shellcode, Worms \\
    \midrule
    \cite{bib15}    & IoT network-trafﬁc dataset   & Blackhole Attack, Opportunistic Service Attack, Distributed Denial-of-Service (DDoS) Attack, Sinkhole Attack, Wormhole Attack\\
    \midrule
    \cite{bib17} & NSL-KDD dataset & protocol used, source 
address, destination address, the time-stamp, services, 
flag, number of failed logins, number of logins, etc  \\
    \midrule
    \cite{Basati2023} & UNSW-NB15 & Reconnaissance, Backdoor, Dos, Exploit, Analysis, Fuzzers, Worms, Shellcode, Generic \\ \\

          & CICIDS2017 & DoS, PortScan, Inﬁltration, Web Attack   \\ \\
     
         & KDDCup99 & DoS, Probe, R2L, U2R \\
     \midrule
    \cite{kamalov} & CSE-CIC-IDS2018 & DDoS \\
     \midrule
    \cite{bib18} & KDDCup 99 & DoS, Probe, R2L, U2R \\ \\
     
     & NSL-KDD & DoS, Probe, R2L, U2R \\ \\
     & UNSW-NB15 & Reconnaissance, Backdoor, Dos, Exploit, Analysis, Fuzzers, Worms, Shellcode, Generic \\ \\
    
      & WSN-DS & Blackhole, Grayhole , Flooding, Scheduling \\ \\
     
      & CICIDS 2017 & SSH-Patator, FTP-Patator, DoS, Web, Bot, DDoS, PortScan \\ \\
     
     & Kyoto & Not specified\\ \\

     & ADFA-LD & Adduser, Java-Meterpreter, Hydra-FTP, Hydra-SSH, Meterpreter, Web-Shell \\ \\

          & ADFA-WD & Not Specified \\
     \midrule
    \cite{bib20} & NSL-KDD dataset & DoS, Probe, R2L, U2R \\
     \midrule
    \cite{masum2021} & NSL-KDD dataset& DoS, Probe, R2L, U2R  \\
     \midrule
   \cite{Khanetal2021} & MQTT-IoT-IDS2020 & Bi-flow,Uni-flow
    Packet-ﬂow\\ \\
    & MQTT dataset& Man in the Middle (MitM), Intrusion in the network, Denial of Services (DoS)\\
     \midrule
    \cite{Gulia2023} & NSL-KDD dataset & DoS, Probe, R2L, U2R\\
     &UNSW-NB15 dataset & Backdoor, Dos, Exploit, Worms\\
     \midrule
     \cite{Syariful2022} & UNSW Bot-IoT dataset & DoS, Theft, and Reconnaissance attacks\\
    \midrule
    \cite{Ashwaq2022} & NSL-KDD dataset & DoS, Probe, R2L, U2R\\
    \midrule
    \cite{aldhaheri_deepdca_2020} & IoT-Bot dataset &  Information Theft, Reconnaissance, DDoS, DoS \\
    \midrule
    \cite{bib39} & KDD CUP99 dataset & DoS, Probe, R2L, U2R \\
    \midrule
    \cite{mohammed2020multilayer} & NSL-KDD dataset &  DoS, Probe, R2L, U2R \\
    \midrule
   \cite{Rosay2022} & CIC-IDS2017 data set & Bot, DDoS, DoS GoldenEye, DoS Hulk, DoS Slowhttptest, DoS Slowloris, FTP-PATATOR, Heartbleed,Infiltration,PortScan,SSH-PATATOR, WebAttack BruteForce, WebAttack SQL Injection, WebAttack XSS\\ \\
 & CSE-CIC-IDS2018 dataset & Bot,BruteForce-Web,BruteForce-XSS, DDoS attack-HOIC, DDoS attack-LOIC-UDP, DDoS attack-LOIC-HTTP, DoS attacks-GoldenEye, DoS attacks-Hulk, DoS attacks-Slowhttptest, DoS attacks-Slowloris, FTP-BruteForce, Infiltration, SQL Injection, SSH-BruteForce\\
         
    \midrule
    \cite{bib13} & BoT-IoT & Information
Theft, Reconnaissance, DDoS, DoS
 \\
     \midrule
    \cite{bib19} & IoT-23 & Malware-1-1,Malware-3-1,Honeypot-4-1,Honeypot-5-1,Honeypot-7-1,Malware-34-1,Malware-43-1\\ \\
        &LITNET-2020 & Smurf ,ICMP-ﬂood,UDP-ﬂood ,TCP SYN-ﬂood,HTTP-ﬂood,LANDattack,Blaster worm,Code red worm,Spam bot’s detection,Reaper worm,Scanning/spread,Packet fragmentation attack  \\ \\
        &NetML-2020 & Bening,Artemis,BitCoinMiner,CCleaner ,Cobalt,Downware,Dridex, Emotet,HTBot,MagicHound,MinerTrojan,PUA,Ramnit,
        Sality,Tinba,TrickBot,Trickster,TrojanDownloader
,Ursnif,WebCompanion\ \\
         \midrule
    \cite{Pooja2021} & NSL-KDD &  DoS, Probe, R2L, U2R\\ \\
     \midrule
    \cite{Alkahtani2021} & IoTID20 dataset & DoS,Mirai ack flooding, Mirai HTTP flooding,Mirai UDP flooding,Scan port OS,Mirai brute force,Scan host port,MITM\\
     \midrule
    \cite{ASHIKU2021239} & UNSW-NB15 dataset& Analysis, Backdoors, DoS, Exploits, Fuzzers,Generic, Recon, Shellcode, Worms \\
     \midrule
    \cite{bib21} & NSL-KDD dataset & DoS, Probe, R2L, U2R \\
     \midrule
    \cite{Safi2022} & Combined DDoS & port scan, DDOSIM, goldeneye, hulk, slowloris,rudy, slowread, SlowHTTPTest, or LOIC types of DDoS attack\\ \\
        & Car hack 2020 & DDoS, Fuzzy, Spoof\\
    \midrule
    \cite{samutha2022} & UNSW-NB18 datasets & Reconnaissance, Backdoor, Dos, Exploit, Analysis, Fuzzers, Worms, Shellcode, Generic\\
    \midrule
    \cite{Gautam2022} & CICIDS2017 dataset &Bot, DDoS, DoS GoldenEye, DoS Hulk, DoS Slowhttptest, DoS Slowloris, FTP-PATATOR, Heartbleed, Infiltration, PortScan, SSH-PATATOR, WebAttack BruteForce, WebAttack SQL Injection, WebAttack XSS \\
    \midrule
    \cite{KASONGO2023} & NSL-KDD dataset& DoS, Probe, R2L, U2R  \\ \\
    & UNSW-NB15 & Reconnaissance, Backdoor, Dos, Exploit, Analysis, Fuzzers, Worms, Shellcode, Generic  \\
    \hline
\end{longtable}
\pagebreak
\section{Discussion}

The thorough examination of intrusion detection systems (IDS) in the attached article highlights an important trend: the sparse use of IDS in the context of the Internet of Things (IoT). There is a clear gap in the application of intrusion detection systems (IDS) that are specifically designed for Internet of Things (IoT) contexts, despite the fact that cyber threats are constantly changing and networked environments are becoming more sophisticated. Convolutional neural networks (CNNs), deep belief networks (DBNs), autoencoders (AE), deep neural networks (DNNs), recurrent neural networks (RNNs), and self-normalizing neural networks (SNNs) are among the many deep learning models that are widely discussed and used in a variety of contexts, including cloud computing and targeted attack scenarios like distributed denial of service (DDoS). However, there seems to be less use of these models in the Internet of Things (IoT) space.\par
In order to improve the efficiency and accuracy of intrusion detection, researchers have shown ingenuity in fusing these deep learning models with cutting-edge algorithms such as Dendritic Cell Algorithm (DCA), Spider Monkey Optimization (SMO), and Group-Artificial Bee Colony (G-ABC). Nonetheless, the results highlight how important it is to pay more attention to and investigate IDS designed for IoT applications. The underrepresentation of IDS in IoT contexts raises concerns about a potential gap in addressing the particular issues presented by IoT networks, despite the abundance of research concentrated on other domains. This observation points to a need for additional study and development to strengthen IoT environments against new cyberthreats and to advance intrusion detection technologies in general inside the quickly growing network of linked devices.

\backmatter

\section{Conclusion}

In this systematic study, we were able to analyze deep learning algorithms that have been used to develop IDS for the last 5 years i.e 2019-2023. The research aimed to examine the current algorithms utilized to create intrusion detection systems and explain how they have been applied to guarantee everyone's safety in cyberspace. Some of the key things that were considered include the datasets used to come up with the algorithms, what Deep learning algorithms have been used in IDS, the attacks covered in the datasets, and the problems that the models developed intended to solve.\par
In future work, it would be of great help to ensure that more studies have been done to come up with better IDS models that can be utilized in the rapidly growing IoT field to prevent advanced cyber threats.\par
\backmatter

\bibliography{sn-bibliography}


\begin{thebibliography}{44}
\ifx \bisbn   \undefined \def \bisbn  #1{ISBN #1}\fi
\ifx \binits  \undefined \def \binits#1{#1}\fi
\ifx \bauthor  \undefined \def \bauthor#1{#1}\fi
\ifx \batitle  \undefined \def \batitle#1{#1}\fi
\ifx \bjtitle  \undefined \def \bjtitle#1{#1}\fi
\ifx \bvolume  \undefined \def \bvolume#1{\textbf{#1}}\fi
\ifx \byear  \undefined \def \byear#1{#1}\fi
\ifx \bissue  \undefined \def \bissue#1{#1}\fi
\ifx \bfpage  \undefined \def \bfpage#1{#1}\fi
\ifx \blpage  \undefined \def \blpage #1{#1}\fi
\ifx \burl  \undefined \def \burl#1{\textsf{#1}}\fi
\ifx \doiurl  \undefined \def \doiurl#1{\url{https://doi.org/#1}}\fi
\ifx \betal  \undefined \def \betal{\textit{et al.}}\fi
\ifx \binstitute  \undefined \def \binstitute#1{#1}\fi
\ifx \binstitutionaled  \undefined \def \binstitutionaled#1{#1}\fi
\ifx \bctitle  \undefined \def \bctitle#1{#1}\fi
\ifx \beditor  \undefined \def \beditor#1{#1}\fi
\ifx \bpublisher  \undefined \def \bpublisher#1{#1}\fi
\ifx \bbtitle  \undefined \def \bbtitle#1{#1}\fi
\ifx \bedition  \undefined \def \bedition#1{#1}\fi
\ifx \bseriesno  \undefined \def \bseriesno#1{#1}\fi
\ifx \blocation  \undefined \def \blocation#1{#1}\fi
\ifx \bsertitle  \undefined \def \bsertitle#1{#1}\fi
\ifx \bsnm \undefined \def \bsnm#1{#1}\fi
\ifx \bsuffix \undefined \def \bsuffix#1{#1}\fi
\ifx \bparticle \undefined \def \bparticle#1{#1}\fi
\ifx \barticle \undefined \def \barticle#1{#1}\fi
\bibcommenthead
\ifx \bconfdate \undefined \def \bconfdate #1{#1}\fi
\ifx \botherref \undefined \def \botherref #1{#1}\fi
\ifx \url \undefined \def \url#1{\textsf{#1}}\fi
\ifx \bchapter \undefined \def \bchapter#1{#1}\fi
\ifx \bbook \undefined \def \bbook#1{#1}\fi
\ifx \bcomment \undefined \def \bcomment#1{#1}\fi
\ifx \oauthor \undefined \def \oauthor#1{#1}\fi
\ifx \citeauthoryear \undefined \def \citeauthoryear#1{#1}\fi
\ifx \endbibitem  \undefined \def \endbibitem {}\fi
\ifx \bconflocation  \undefined \def \bconflocation#1{#1}\fi
\ifx \arxivurl  \undefined \def \arxivurl#1{\textsf{#1}}\fi
\csname PreBibitemsHook\endcsname

\bibitem[\protect\citeauthoryear{frontpor}{2019}]{bib1}
\begin{botherref}
\oauthor{\bsnm{frontpor}}:
Explore the {Importance} of {Computer} {Networking} {\textbar} {Warwick}
(2019).
\url{https://www.warwickinc.com/blog/the-importance-of-computer-networking/}
Accessed 2023-12-13
\end{botherref}
\endbibitem

\bibitem[\protect\citeauthoryear{Abraham and Bindu}{2021}]{bib2}
\begin{bchapter}
\bauthor{\bsnm{Abraham}, \binits{J.A.}},
\bauthor{\bsnm{Bindu}, \binits{V.R.}}:
\bctitle{Intrusion detection and prevention in networks using machine learning and deep learning approaches: A review}.
In: \bbtitle{2021 International Conference on Advancements in Electrical, Electronics, Communication, Computing and Automation (ICAECA)},
pp. \bfpage{1}--\blpage{4}
(\byear{2021}).
\doiurl{10.1109/ICAECA52838.2021.9675595}
\end{bchapter}
\endbibitem

\bibitem[\protect\citeauthoryear{Song}{2022}]{bib3}
\begin{barticle}
\bauthor{\bsnm{Song}, \binits{J.}}:
\batitle{Preschool cyber security management system based on intelligent agents}.
\bjtitle{Computational Intelligence and Neuroscience}
\bvolume{2022},
\bfpage{1992429}
(\byear{2022})
\doiurl{10.1155/2022/1992429}
\end{barticle}
\endbibitem

\bibitem[\protect\citeauthoryear{Momand et~al.}{2023}]{bib4}
\begin{barticle}
\bauthor{\bsnm{Momand}, \binits{A.}},
\bauthor{\bsnm{Jan}, \binits{S.U.}},
\bauthor{\bsnm{Ramzan}, \binits{N.}}:
\batitle{A {Systematic} and {Comprehensive} {Survey} of {Recent} {Advances} in {Intrusion} {Detection} {Systems} {Using} {Machine} {Learning}: {Deep} {Learning}, {Datasets}, and {Attack} {Taxonomy}}.
\bjtitle{Journal of Sensors}
\bvolume{2023},
\bfpage{6048087}
(\byear{2023})
\doiurl{10.1155/2023/6048087} .
\bcomment{Publisher: Hindawi}
\end{barticle}
\endbibitem

\bibitem[\protect\citeauthoryear{Vinayakumar et~al.}{2019}]{bib5}
\begin{barticle}
\bauthor{\bsnm{Vinayakumar}, \binits{R.}},
\bauthor{\bsnm{Alazab}, \binits{M.}},
\bauthor{\bsnm{Soman}, \binits{K.P.}},
\bauthor{\bsnm{Poornachandran}, \binits{P.}},
\bauthor{\bsnm{Al-Nemrat}, \binits{A.}},
\bauthor{\bsnm{Venkatraman}, \binits{S.}}:
\batitle{Deep learning approach for intelligent intrusion detection system}.
\bjtitle{IEEE Access}
\bvolume{7},
\bfpage{41525}--\blpage{41550}
(\byear{2019})
\doiurl{10.1109/ACCESS.2019.2895334}
\end{barticle}
\endbibitem

\bibitem[\protect\citeauthoryear{Lansky et~al.}{2021}]{bib6}
\begin{barticle}
\bauthor{\bsnm{Lansky}, \binits{J.}},
\bauthor{\bsnm{Ali}, \binits{S.}},
\bauthor{\bsnm{Mohammadi}, \binits{M.}},
\bauthor{\bsnm{Majeed}, \binits{M.K.}},
\bauthor{\bsnm{Karim}, \binits{S.H.T.}},
\bauthor{\bsnm{Rashidi}, \binits{S.}},
\bauthor{\bsnm{Hosseinzadeh}, \binits{M.}},
\bauthor{\bsnm{Rahmani}, \binits{A.M.}}:
\batitle{Deep learning-based intrusion detection systems: A systematic review}.
\bjtitle{IEEE Access}
\bvolume{9},
\bfpage{101574}--\blpage{101599}
(\byear{2021})
\doiurl{10.1109/ACCESS.2021.3097247}
\end{barticle}
\endbibitem

\bibitem[\protect\citeauthoryear{Liu and Lang}{2019}]{bib7}
\begin{botherref}
\oauthor{\bsnm{Liu}, \binits{H.}},
\oauthor{\bsnm{Lang}, \binits{B.}}:
Machine learning and deep learning methods for intrusion detection systems: A survey.
Applied Sciences
\textbf{9}(20)
(2019)
\doiurl{10.3390/app9204396}
\end{botherref}
\endbibitem

\bibitem[\protect\citeauthoryear{Jakhar and Kaur}{2020}]{bib8}
\begin{barticle}
\bauthor{\bsnm{Jakhar}, \binits{D.}},
\bauthor{\bsnm{Kaur}, \binits{I.}}:
\batitle{{Artificial intelligence, machine learning and deep learning: definitions and differences}}.
\bjtitle{Clinical and Experimental Dermatology}
\bvolume{45}(\bissue{1}),
\bfpage{131}--\blpage{132}
(\byear{2020})
\doiurl{10.1111/ced.14029}
{\href{https://arxiv.org/abs/https://academic.oup.com/ced/article-pdf/45/1/131/46920173/ced0131.pdf}{{https://academic.oup.com/ced/article-pdf/45/1/131/46920173/ced0131.pdf}}}
\end{barticle}
\endbibitem

\bibitem[\protect\citeauthoryear{Deng and Yu}{2014}]{bib9}
\begin{barticle}
\bauthor{\bsnm{Deng}, \binits{L.}},
\bauthor{\bsnm{Yu}, \binits{D.}}:
\batitle{Deep learning: Methods and applications}.
\bjtitle{Foundations and Trends® in Signal Processing}
\bvolume{7}(\bissue{3–4}),
\bfpage{197}--\blpage{387}
(\byear{2014})
\doiurl{10.1561/2000000039}
\end{barticle}
\endbibitem

\bibitem[\protect\citeauthoryear{Biswal}{2023}]{bib10}
\begin{botherref}
\oauthor{\bsnm{Biswal}, \binits{A.}}:
Top 10 {Deep} {Learning} {Algorithms} {You} {Should} {Know} in 2023
(2023).
\url{https://www.simplilearn.com/tutorials/deep-learning-tutorial/deep-learning-algorithm}
Accessed 2023-12-22
\end{botherref}
\endbibitem

\bibitem[\protect\citeauthoryear{Khan et~al.}{2019}]{bib11}
\begin{bchapter}
\bauthor{\bsnm{Khan}, \binits{R.U.}},
\bauthor{\bsnm{Zhang}, \binits{X.}},
\bauthor{\bsnm{Alazab}, \binits{M.}},
\bauthor{\bsnm{Kumar}, \binits{R.}}:
\bctitle{An improved convolutional neural network model for intrusion detection in networks}.
In: \bbtitle{2019 Cybersecurity and Cyberforensics Conference (CCC)},
pp. \bfpage{74}--\blpage{77}
(\byear{2019}).
\doiurl{10.1109/CCC.2019.000-6}
\end{bchapter}
\endbibitem

\bibitem[\protect\citeauthoryear{Liu et~al.}{2019}]{bib12}
\begin{bchapter}
\bauthor{\bsnm{Liu}, \binits{W.}},
\bauthor{\bsnm{Liu}, \binits{X.}},
\bauthor{\bsnm{Di}, \binits{X.}},
\bauthor{\bsnm{Qi}, \binits{H.}}:
\bctitle{A novel network intrusion detection algorithm based on fast fourier transformation}.
In: \bbtitle{2019 1st International Conference on Industrial Artificial Intelligence (IAI)},
pp. \bfpage{1}--\blpage{6}
(\byear{2019}).
\doiurl{10.1109/ICIAI.2019.8850770}
\end{bchapter}
\endbibitem

\bibitem[\protect\citeauthoryear{Hu et~al.}{2021}]{Hu2021}
\begin{barticle}
\bauthor{\bsnm{Hu}, \binits{Y.}},
\bauthor{\bsnm{Bai}, \binits{F.}},
\bauthor{\bsnm{Yang}, \binits{X.}},
\bauthor{\bsnm{Liu}, \binits{Y.}}:
\batitle{Idsdl: a sensitive intrusion detection system based on deep learning}.
\bjtitle{EURASIP Journal on Wireless Communications and Networking}
\bvolume{2021}(\bissue{1}),
\bfpage{95}
(\byear{2021})
\doiurl{10.1186/s13638-021-01900-y}
\end{barticle}
\endbibitem

\bibitem[\protect\citeauthoryear{Kumar~Silivery et~al.}{2023}]{Arun2023}
\begin{bchapter}
\bauthor{\bsnm{Kumar~Silivery}, \binits{A.}},
\bauthor{\bsnm{Mohan~Rao}, \binits{K.R.}},
\bauthor{\bsnm{Solleti}, \binits{R.}}:
\bctitle{An advanced intrusion detection algorithm for network traffic using convolution neural network}.
In: \bbtitle{2023 Fifth International Conference on Electrical, Computer and Communication Technologies (ICECCT)},
pp. \bfpage{01}--\blpage{05}
(\byear{2023}).
\doiurl{10.1109/ICECCT56650.2023.10179767}
\end{bchapter}
\endbibitem

\bibitem[\protect\citeauthoryear{A.~Alissa et~al.}{2022}]{Khalid2022}
\begin{botherref}
\oauthor{\bsnm{A.~Alissa}, \binits{K.}},
\oauthor{\bsnm{S.~Alrayes}, \binits{F.}},
\oauthor{\bsnm{Tarmissi}, \binits{K.}},
\oauthor{\bsnm{Yafoz}, \binits{A.}},
\oauthor{\bsnm{Alsini}, \binits{R.}},
\oauthor{\bsnm{Alghushairy}, \binits{O.}},
\oauthor{\bsnm{Othman}, \binits{M.}},
\oauthor{\bsnm{Motwakel}, \binits{A.}}:
Planet optimization with deep convolutional neural network for lightweight intrusion detection in resource-constrained iot networks.
Applied Sciences
\textbf{12}(17)
(2022)
\doiurl{10.3390/app12178676}
\end{botherref}
\endbibitem

\bibitem[\protect\citeauthoryear{Man and Sun}{2021}]{Manandsun2021}
\begin{barticle}
\bauthor{\bsnm{Man}, \binits{J.}},
\bauthor{\bsnm{Sun}, \binits{G.}}:
\batitle{A residual learning-based network intrusion detection system}.
\bjtitle{Security and Communication Networks}
\bvolume{2021},
\bfpage{5593435}
(\byear{2021})
\doiurl{10.1155/2021/5593435}
\end{barticle}
\endbibitem

\bibitem[\protect\citeauthoryear{Yang et~al.}{2019}]{bib14}
\begin{botherref}
\oauthor{\bsnm{Yang}, \binits{Y.}},
\oauthor{\bsnm{Zheng}, \binits{K.}},
\oauthor{\bsnm{Wu}, \binits{C.}},
\oauthor{\bsnm{Niu}, \binits{X.}},
\oauthor{\bsnm{Yang}, \binits{Y.}}:
Building an {Effective} {Intrusion} {Detection} {System} {Using} the {Modified} {Density} {Peak} {Clustering} {Algorithm} and {Deep} {Belief} {Networks}.
Applied Sciences
\textbf{9}(2)
(2019)
\doiurl{10.3390/app9020238}
\end{botherref}
\endbibitem

\bibitem[\protect\citeauthoryear{Thamilarasu and Chawla}{2019}]{bib15}
\begin{botherref}
\oauthor{\bsnm{Thamilarasu}, \binits{G.}},
\oauthor{\bsnm{Chawla}, \binits{S.}}:
Towards deep-learning-driven intrusion detection for the internet of things.
Sensors
\textbf{19}(9)
(2019)
\doiurl{10.3390/s19091977}
\end{botherref}
\endbibitem

\bibitem[\protect\citeauthoryear{Al-Hadhrami and Hussain}{2021}]{bib16}
\begin{barticle}
\bauthor{\bsnm{Al-Hadhrami}, \binits{Y.}},
\bauthor{\bsnm{Hussain}, \binits{F.K.}}:
\batitle{{DDoS} attacks in {IoT} networks: a comprehensive systematic literature review}.
\bjtitle{World Wide Web}
\bvolume{24}(\bissue{3}),
\bfpage{971}--\blpage{1001}
(\byear{2021})
\doiurl{10.1007/s11280-020-00855-2}
\end{barticle}
\endbibitem

\bibitem[\protect\citeauthoryear{Gurung et~al.}{2019}]{bib17}
\begin{barticle}
\bauthor{\bsnm{Gurung}, \binits{S.}},
\bauthor{\bsnm{Ghose}, \binits{M.K.}},
\bauthor{\bsnm{Subedi}, \binits{A.}}:
\batitle{Deep learning approach on network intrusion detection system using {NSL}-{KDD} dataset}.
\bjtitle{International Journal of Computer Network and Information Security}
\bvolume{11}(\bissue{3}),
\bfpage{8}--\blpage{14}
(\byear{2019})
\doiurl{10.5815/ijcnis.2019.03.02} .
\bcomment{Publisher: Modern Education and Computer Science Press}
\end{barticle}
\endbibitem

\bibitem[\protect\citeauthoryear{Telikani and Gandomi}{2021}]{TELIKANI2021100122}
\begin{barticle}
\bauthor{\bsnm{Telikani}, \binits{A.}},
\bauthor{\bsnm{Gandomi}, \binits{A.H.}}:
\batitle{Cost-sensitive stacked auto-encoders for intrusion detection in the internet of things}.
\bjtitle{Internet of Things}
\bvolume{14},
\bfpage{100122}
(\byear{2021})
\doiurl{10.1016/j.iot.2019.100122}
\end{barticle}
\endbibitem

\bibitem[\protect\citeauthoryear{Basati and Faghih}{2023}]{Basati2023}
\begin{barticle}
\bauthor{\bsnm{Basati}, \binits{A.}},
\bauthor{\bsnm{Faghih}, \binits{M.M.}}:
\batitle{Apae: an iot intrusion detection system using asymmetric parallel auto-encoder}.
\bjtitle{Neural Computing and Applications}
\bvolume{35}(\bissue{7}),
\bfpage{4813}--\blpage{4833}
(\byear{2023})
\doiurl{10.1007/s00521-021-06011-9}
\end{barticle}
\endbibitem

\bibitem[\protect\citeauthoryear{Kamalov et~al.}{2021}]{kamalov}
\begin{bchapter}
\bauthor{\bsnm{Kamalov}, \binits{F.}},
\bauthor{\bsnm{Zgheib}, \binits{R.}},
\bauthor{\bsnm{Leung}, \binits{H.H.}},
\bauthor{\bsnm{Al-Gindy}, \binits{A.}},
\bauthor{\bsnm{Moussa}, \binits{S.}}:
\bctitle{Autoencoder-based intrusion detection system}.
In: \bbtitle{2021 International Conference on Engineering and Emerging Technologies (ICEET)},
pp. \bfpage{1}--\blpage{5}
(\byear{2021}).
\doiurl{10.1109/ICEET53442.2021.9659562}
\end{bchapter}
\endbibitem

\bibitem[\protect\citeauthoryear{Vinayakumar et~al.}{2019}]{bib18}
\begin{barticle}
\bauthor{\bsnm{Vinayakumar}, \binits{R.}},
\bauthor{\bsnm{Alazab}, \binits{M.}},
\bauthor{\bsnm{Soman}, \binits{K.P.}},
\bauthor{\bsnm{Poornachandran}, \binits{P.}},
\bauthor{\bsnm{Al-Nemrat}, \binits{A.}},
\bauthor{\bsnm{Venkatraman}, \binits{S.}}:
\batitle{Deep {Learning} {Approach} for {Intelligent} {Intrusion} {Detection} {System}}.
\bjtitle{IEEE Access}
\bvolume{7},
\bfpage{41525}--\blpage{41550}
(\byear{2019})
\doiurl{10.1109/ACCESS.2019.2895334}
\end{barticle}
\endbibitem

\bibitem[\protect\citeauthoryear{Khare et~al.}{2020}]{bib20}
\begin{botherref}
\oauthor{\bsnm{Khare}, \binits{N.}},
\oauthor{\bsnm{Devan}, \binits{P.}},
\oauthor{\bsnm{Chowdhary}, \binits{C.L.}},
\oauthor{\bsnm{Bhattacharya}, \binits{S.}},
\oauthor{\bsnm{Singh}, \binits{G.}},
\oauthor{\bsnm{Singh}, \binits{S.}},
\oauthor{\bsnm{Yoon}, \binits{B.}}:
{SMO}-{DNN}: {Spider} {Monkey} {Optimization} and {Deep} {Neural} {Network} {Hybrid} {Classifier} {Model} for {Intrusion} {Detection}.
Electronics
\textbf{9}(4)
(2020)
\doiurl{10.3390/electronics9040692}
\end{botherref}
\endbibitem

\bibitem[\protect\citeauthoryear{Masum and Shahriar}{2020}]{masum2021}
\begin{bchapter}
\bauthor{\bsnm{Masum}, \binits{M.}},
\bauthor{\bsnm{Shahriar}, \binits{H.}}:
\bctitle{Tl-nid: Deep neural network with transfer learning for network intrusion detection}.
In: \bbtitle{2020 15th International Conference for Internet Technology and Secured Transactions (ICITST)},
pp. \bfpage{1}--\blpage{7}
(\byear{2020}).
\doiurl{10.23919/ICITST51030.2020.9351317}
\end{bchapter}
\endbibitem

\bibitem[\protect\citeauthoryear{Khan et~al.}{2021}]{Khanetal2021}
\begin{botherref}
\oauthor{\bsnm{Khan}, \binits{M.A.}},
\oauthor{\bsnm{Khan}, \binits{M.A.}},
\oauthor{\bsnm{Jan}, \binits{S.U.}},
\oauthor{\bsnm{Ahmad}, \binits{J.}},
\oauthor{\bsnm{Jamal}, \binits{S.S.}},
\oauthor{\bsnm{Shah}, \binits{A.A.}},
\oauthor{\bsnm{Pitropakis}, \binits{N.}},
\oauthor{\bsnm{Buchanan}, \binits{W.J.}}:
A deep learning-based intrusion detection system for mqtt enabled iot.
Sensors
\textbf{21}(21)
(2021)
\doiurl{10.3390/s21217016}
\end{botherref}
\endbibitem

\bibitem[\protect\citeauthoryear{Gulia et~al.}{2023}]{Gulia2023}
\begin{barticle}
\bauthor{\bsnm{Gulia}, \binits{N.}},
\bauthor{\bsnm{Solanki}, \binits{K.}},
\bauthor{\bsnm{Dalal}, \binits{S.}},
\bauthor{\bsnm{Dhankhar}, \binits{A.}},
\bauthor{\bsnm{Dahiya}, \binits{O.}},
\bauthor{\bsnm{Salmaan}, \binits{N.U.}}:
\batitle{Intrusion detection system using the g-abc with deep neural network in cloud environment}.
\bjtitle{Scientific Programming}
\bvolume{2023},
\bfpage{7210034}
(\byear{2023})
\doiurl{10.1155/2023/7210034}
\end{barticle}
\endbibitem

\bibitem[\protect\citeauthoryear{Ikhwan et~al.}{2022}]{Syariful2022}
\begin{bchapter}
\bauthor{\bsnm{Ikhwan}, \binits{S.}},
\bauthor{\bsnm{Wibowo}, \binits{A.}},
\bauthor{\bsnm{Warsito}, \binits{B.}}:
\bctitle{Intrusion detection using deep neural network algorithm on the internet of things}.
In: \bbtitle{2022 IEEE International Conference on Communication, Networks and Satellite (COMNETSAT)},
pp. \bfpage{84}--\blpage{87}
(\byear{2022}).
\doiurl{10.1109/COMNETSAT56033.2022.9994499}
\end{bchapter}
\endbibitem

\bibitem[\protect\citeauthoryear{Almutairi and Abdulghani~Alshargabi}{2022}]{Ashwaq2022}
\begin{bchapter}
\bauthor{\bsnm{Almutairi}, \binits{A.F.}},
\bauthor{\bsnm{Abdulghani~Alshargabi}, \binits{A.}}:
\bctitle{Using deep learning technique to protect internet network from intrusion in iot environment}.
In: \bbtitle{2022 2nd International Conference on Emerging Smart Technologies and Applications (eSmarTA)},
pp. \bfpage{1}--\blpage{6}
(\byear{2022}).
\doiurl{10.1109/eSmarTA56775.2022.9935467}
\end{bchapter}
\endbibitem

\bibitem[\protect\citeauthoryear{Aldhaheri et~al.}{2020}]{aldhaheri_deepdca_2020}
\begin{botherref}
\oauthor{\bsnm{Aldhaheri}, \binits{S.}},
\oauthor{\bsnm{Alghazzawi}, \binits{D.}},
\oauthor{\bsnm{Cheng}, \binits{L.}},
\oauthor{\bsnm{Alzahrani}, \binits{B.}},
\oauthor{\bsnm{Al-Barakati}, \binits{A.}}:
{DeepDCA}: {Novel} {Network}-{Based} {Detection} of {IoT} {Attacks} {Using} {Artificial} {Immune} {System}.
Applied Sciences
\textbf{10}(6)
(2020)
\doiurl{10.3390/app10061909}
\end{botherref}
\endbibitem

\bibitem[\protect\citeauthoryear{Louati and Ktata}{2020}]{bib39}
\begin{barticle}
\bauthor{\bsnm{Louati}, \binits{F.}},
\bauthor{\bsnm{Ktata}, \binits{F.B.}}:
\batitle{A deep learning-based multi-agent system for intrusion detection}.
\bjtitle{SN Applied Sciences}
\bvolume{2}(\bissue{4}),
\bfpage{675}
(\byear{2020})
\doiurl{10.1007/s42452-020-2414-z}
\end{barticle}
\endbibitem

\bibitem[\protect\citeauthoryear{Mohammed et~al.}{2020}]{mohammed2020multilayer}
\begin{barticle}
\bauthor{\bsnm{Mohammed}, \binits{A.J.}},
\bauthor{\bsnm{Arif}, \binits{M.H.}},
\bauthor{\bsnm{Ali}, \binits{A.A.}}:
\batitle{A multilayer perceptron artificial neural network approach for improving the accuracy of intrusion detection systems}.
\bjtitle{IAES International Journal of Artificial Intelligence}
\bvolume{9}(\bissue{4}),
\bfpage{609}
(\byear{2020})
\doiurl{10.11591/ijai.v9.i4.pp609-615}
\end{barticle}
\endbibitem

\bibitem[\protect\citeauthoryear{Rosay et~al.}{2022}]{Rosay2022}
\begin{barticle}
\bauthor{\bsnm{Rosay}, \binits{A.}},
\bauthor{\bsnm{Riou}, \binits{K.}},
\bauthor{\bsnm{Carlier}, \binits{F.}},
\bauthor{\bsnm{Leroux}, \binits{P.}}:
\batitle{Multi-layer perceptron for network intrusion detection}.
\bjtitle{Annals of Telecommunications}
\bvolume{77}(\bissue{5}),
\bfpage{371}--\blpage{394}
(\byear{2022})
\doiurl{10.1007/s12243-021-00852-0}
\end{barticle}
\endbibitem

\bibitem[\protect\citeauthoryear{Susilo and Sari}{2020}]{bib13}
\begin{botherref}
\oauthor{\bsnm{Susilo}, \binits{B.}},
\oauthor{\bsnm{Sari}, \binits{R.F.}}:
Intrusion detection in iot networks using deep learning algorithm.
Information
\textbf{11}(5)
(2020)
\doiurl{10.3390/info11050279}
\end{botherref}
\endbibitem

\bibitem[\protect\citeauthoryear{Dutta et~al.}{2020}]{bib19}
\begin{botherref}
\oauthor{\bsnm{Dutta}, \binits{V.}},
\oauthor{\bsnm{Choraś}, \binits{M.}},
\oauthor{\bsnm{Pawlicki}, \binits{M.}},
\oauthor{\bsnm{Kozik}, \binits{R.}}:
A {Deep} {Learning} {Ensemble} for {Network} {Anomaly} and {Cyber}-{Attack} {Detection}.
Sensors
\textbf{20}(16)
(2020)
\doiurl{10.3390/s20164583}
\end{botherref}
\endbibitem

\bibitem[\protect\citeauthoryear{Shettar et~al.}{2021}]{Pooja2021}
\begin{bchapter}
\bauthor{\bsnm{Shettar}, \binits{P.}},
\bauthor{\bsnm{Kachavimath}, \binits{A.V.}},
\bauthor{\bsnm{Mulla}, \binits{M.M.}},
\bauthor{\bsnm{G}, \binits{N.D.}},
\bauthor{\bsnm{Hanchinmani}, \binits{G.}}:
\bctitle{Intrusion detection system using mlp and chaotic neural networks}.
In: \bbtitle{2021 International Conference on Computer Communication and Informatics (ICCCI)},
pp. \bfpage{1}--\blpage{4}
(\byear{2021}).
\doiurl{10.1109/ICCCI50826.2021.9457024}
\end{bchapter}
\endbibitem

\bibitem[\protect\citeauthoryear{Alkahtani and Aldhyani}{2021}]{Alkahtani2021}
\begin{barticle}
\bauthor{\bsnm{Alkahtani}, \binits{H.}},
\bauthor{\bsnm{Aldhyani}, \binits{T.H.H.}}:
\batitle{Intrusion detection system to advance internet of things infrastructure-based deep learning algorithms}.
\bjtitle{Complexity}
\bvolume{2021},
\bfpage{5579851}
(\byear{2021})
\doiurl{10.1155/2021/5579851}
\end{barticle}
\endbibitem

\bibitem[\protect\citeauthoryear{Ashiku and Dagli}{2021}]{ASHIKU2021239}
\begin{barticle}
\bauthor{\bsnm{Ashiku}, \binits{L.}},
\bauthor{\bsnm{Dagli}, \binits{C.}}:
\batitle{Network intrusion detection system using deep learning}.
\bjtitle{Procedia Computer Science}
\bvolume{185},
\bfpage{239}--\blpage{247}
(\byear{2021})
\doiurl{10.1016/j.procs.2021.05.025} .
\bcomment{Big Data, IoT, and AI for a Smarter Future}
\end{barticle}
\endbibitem

\bibitem[\protect\citeauthoryear{Muhuri et~al.}{2020}]{bib21}
\begin{botherref}
\oauthor{\bsnm{Muhuri}, \binits{P.S.}},
\oauthor{\bsnm{Chatterjee}, \binits{P.}},
\oauthor{\bsnm{Yuan}, \binits{X.}},
\oauthor{\bsnm{Roy}, \binits{K.}},
\oauthor{\bsnm{Esterline}, \binits{A.}}:
Using a {Long} {Short}-{Term} {Memory} {Recurrent} {Neural} {Network} ({LSTM}-{RNN}) to {Classify} {Network} {Attacks}.
Information
\textbf{11}(5)
(2020)
\doiurl{10.3390/info11050243}
\end{botherref}
\endbibitem

\bibitem[\protect\citeauthoryear{Ullah et~al.}{2022}]{Safi2022}
\begin{botherref}
\oauthor{\bsnm{Ullah}, \binits{S.}},
\oauthor{\bsnm{Khan}, \binits{M.A.}},
\oauthor{\bsnm{Ahmad}, \binits{J.}},
\oauthor{\bsnm{Jamal}, \binits{S.S.}},
\oauthor{\bsnm{Huma}, \binits{Z.}},
\oauthor{\bsnm{Hassan}, \binits{M.T.}},
\oauthor{\bsnm{Pitropakis}, \binits{N.}},
\oauthor{\bsnm{Arshad}},
\oauthor{\bsnm{Buchanan}, \binits{W.J.}}:
Hdl-ids: A hybrid deep learning architecture for intrusion detection in the internet of vehicles.
Sensors
\textbf{22}(4)
(2022)
\doiurl{10.3390/s22041340}
\end{botherref}
\endbibitem

\bibitem[\protect\citeauthoryear{Amutha et~al.}{2022}]{samutha2022}
\begin{bchapter}
\bauthor{\bsnm{Amutha}, \binits{S.}},
\bauthor{\bsnm{R}, \binits{K.}},
\bauthor{\bsnm{R}, \binits{S.}},
\bauthor{\bsnm{M}, \binits{K.}}:
\bctitle{Secure network intrusion detection system using nid-rnn based deep learning}.
In: \bbtitle{2022 International Conference on Advances in Computing, Communication and Applied Informatics (ACCAI)},
pp. \bfpage{1}--\blpage{5}
(\byear{2022}).
\doiurl{10.1109/ACCAI53970.2022.9752526}
\end{bchapter}
\endbibitem

\bibitem[\protect\citeauthoryear{Gautam et~al.}{2022}]{Gautam2022}
\begin{botherref}
\oauthor{\bsnm{Gautam}, \binits{S.}},
\oauthor{\bsnm{Henry}, \binits{A.}},
\oauthor{\bsnm{Zuhair}, \binits{M.}},
\oauthor{\bsnm{Rashid}, \binits{M.}},
\oauthor{\bsnm{Javed}, \binits{A.R.}},
\oauthor{\bsnm{Maddikunta}, \binits{P.K.R.}}:
A composite approach of intrusion detection systems: Hybrid rnn and correlation-based feature optimization.
Electronics
\textbf{11}(21)
(2022)
\doiurl{10.3390/electronics11213529}
\end{botherref}
\endbibitem

\bibitem[\protect\citeauthoryear{Kasongo}{2023}]{KASONGO2023}
\begin{barticle}
\bauthor{\bsnm{Kasongo}, \binits{S.M.}}:
\batitle{A deep learning technique for intrusion detection system using a recurrent neural networks based framework}.
\bjtitle{Computer Communications}
\bvolume{199},
\bfpage{113}--\blpage{125}
(\byear{2023})
\doiurl{10.1016/j.comcom.2022.12.010}
\end{barticle}
\endbibitem

\end{thebibliography}

\end{document}